\documentclass[11pt,a4paper]{article}

\usepackage{fullpage}
\usepackage{amsmath}
\usepackage{amssymb}
\usepackage{mathrsfs}
\usepackage{graphicx}
\usepackage{psfrag}
\usepackage[all,poly,curve,knot,cmtip]{xy}
\usepackage{citesort}

\newcommand {\be} {\begin{eqnarray*}}
\newcommand {\ee} {\end{eqnarray*}}
\newcommand {\bea} {\begin{eqnarray}}
\newcommand {\eea} {\end{eqnarray}}

\newcommand{\bm}[1]{\boldsymbol{#1}}

\newcommand{\ave}[1]{\langle {#1} \rangle}

\newcommand{\tq}{\tilde{q}}
\newcommand{\tomega}{\tilde{\omega}}

\begin{document}

\title{Isolated horizons in higher-dimensional
Einstein-Gauss-Bonnet gravity}
\author{\textbf{Tom$\acute{\mbox{a}}\check{\mbox{s}}$
Liko}\footnote{Electronic mail: tliko@math.mun.ca}\\
\\{\small \it Department of Physics and Physical Oceanography}\\
{\small \it Memorial University of Newfoundland}\\
{\small \it St. John's, Newfoundland, Canada, A1B 3X7}\\
\\\textbf{Ivan Booth}\footnote{Electronic mail: ibooth@math.mun.ca}\\
\\{\small \it Department of Mathematics and Statistics}\\
{\small \it Memorial University of Newfoundland}\\
{\small \it St. John's, Newfoundland, Canada, A1C 5S7}}

\maketitle





\begin{abstract}

The isolated horizon framework was introduced in order
to provide a local description of black holes that are
in equilibrium with their (possibly dynamic) environment.
Over the past several years, the framework has been
extended to include matter fields (dilaton, Yang-Mills
etc) in $D=4$ dimensions and cosmological constant in
$D\geq3$ dimensions.  In this article we present a
further extension of the framework that includes black
holes in higher-dimensional Einstein-Gauss-Bonnet (EGB)
gravity.  In particular, we construct a covariant
phase space for EGB gravity in arbitrary dimensions
which allows us to derive the first law.  We find
that the entropy of a weakly isolated and non-rotating
horizon is given by
\be
\mathcal{S} = \frac{1}{4G_{D}}\oint_{S^{D-2}}
    \bm{\tilde{\epsilon}}(1 + 2\alpha\mathcal{R}) \; .
\ee
In this expression $S^{D-2}$ is the $(D-2)$-dimensional
cross section of the horizon with area form
$\bm{\tilde{\epsilon}}$ and Ricci scalar $\mathcal{R}$,
$G_{D}$ is the $D$-dimensional Newton constant and
$\alpha$ is the Gauss-Bonnet parameter.  This
expression for the horizon entropy is in agreement
with those predicted by the Euclidean and Noether
charge methods.  Thus we extend the isolated horizon
framework beyond Einstein gravity.

\end{abstract}

\hspace{0.3cm}\textbf{PACS}: 04.20.Fy; 04.50.$+$h; 04.70.Bw


\section{Introduction}

The isolated horizon framework \cite{abdfklw,ashkri,booth,goujar}
provides a very elegant mathematical description of the mechanics
of black holes by replacing the event horizon with an inner boundary
contained in the spacetime manifold.  There are several reasons to
use this quasilocal description of black holes in favour of the old
one that was developed in the seventies
\cite{bekenstein1,bekenstein2,bch,hawking1}.  Among the more
significant are the following: (i) The conventional definition of
a black hole is of a non-local nature because the entire future of
the spacetime must be known before the event horizon can be located;
(ii) If a black hole is assumed to be in equilibrium, then the
surrounding spacetime must also be in equilibrium.  This situation
is clearly not realistic, as radiation and other forms of matter
outside the black hole may be dynamical, while only the hole itself
is in equilibrium; and (iii) The conventional definitions of energy
and angular momentum for a black hole are defined in terms of
asymptotic infinity; the first law, for instance, relates quantities
that are defined at spatial infinity to quantities that are defined
at the horizon.  Clearly, then, a more local notion of black holes
must be introduced to take these issues into account.

Isolated horizons provide such a description, by ``imitating'' the
existence of a Killing vector that becomes null at the horizon.  It
turns out that the existence of an expansion-free null normal
\emph{at the horizon} is sufficient for the zeroth and first laws
of black-hole mechanics to be satisfied.  This is the only physical
assumption in the boundary conditions.  In particular, the zeroth
law follows from basic differential geometry, the energy conditions
and the Raychaudhuri equation.  The first law then follows as a
necessary and sufficient condition in the Hamiltonian evolution upon
choosing an appropriate time translation vector field that points in
the direction of the null normal.  However, unlike its predecessor,
the first law for an isolated horizon relates quantities that are
all defined on the horizon.  For example, the first law for rotating
isolated horizons in Einstein-Maxwell (EM) theory states that
variations of the mass $M_{\Delta}$, surface area $a_{\Delta}$,
angular momentum $J_{\Delta}$ and charge $Q_{\Delta}$ are related
via
\bea
\delta M_{\Delta} = \frac{\kappa}{8\pi G}\delta a_{\Delta}
                    + \Omega\delta J_{\Delta} + \Phi\delta Q_{\Delta} \, ;
\eea
the parameters $\kappa$, $\Omega$, $\Phi$ are, respectively, the
surface gravity, angular velocity and electric potential.  This
is the equilibrium form of the first law which relates the changes
between two nearby equilibrium states within the space of all
solutions.

Isolated horizons have been extensively studied in Einstein gravity.
In particular, the canonical phase space and covariant phase space
were constructed first in terms of complex self-dual connections and
$SL(2,\mathbb{C})$ soldering forms \cite{ack,abf,adw}.  Shortly
afterwards followed a detailed study of dilaton couplings and
Yang-Mills fields \cite{ashcor,corsud}.  The formalism was then
refined and polished by re-expressing the covariant phase space in
terms of real Lorentz connections and tetrads \cite{afk}, which paved
the way for extensions to include e.g. rotation \cite{abl1} and
non-minimally coupled scalar fields \cite{acs}.  Geometrical issues
were extensively studied in
\cite{abl2,lewpaw1,boofai1}.  The framework was extended to
higher-dimensional spacetimes in \cite{lewpaw2,klp,apv}.
Important questions that need to be addressed are the following:
\emph{Can the isolated horizon framework be extended beyond
Einstein gravity?}; and, if so, \emph{Can the resulting framework
be extended to include matter couplings?}  The aim of the present
work is to answer the first of these questions in the affirmative,
by extending the framework to Einstein-Gauss-Bonnet (EGB) gravity
in arbitrary dimensions.

\section{Gauss-Bonnet term in the second-order formulation}

The appearance of curvature-squared terms in the effective action
for gravity from superstring theory is well known \cite{chsw}.
This alone is enough justification for studying the effects of
these extra terms on gravitational objects in higher dimensions.
In addition, there are now other physical models of unification
that employ a large extra dimension, including e.g. braneworld
cosmology \cite{maartens} and induced-matter theory
\cite{low}.

In four dimensions, there is a unique combination of higher-curvature
terms containing at most second derivatives of the metric $g_{ab}$
($a,b,\ldots\in\{0,\ldots,3\}$) that can be added to the Einstein-Hilbert
action, such that the equations of motion are the (vacuum) Einstein field
equations.  This is the Gauss-Bonnet (GB) term \cite{myers,padilla}
\bea
\mathcal{L}_{GB} = R^{2} - 4R_{ab}R^{ab} + R_{abcd}R^{abcd} \, ,
\eea
where $R_{abcd}$ is the Riemann curvature tensor,
$R_{ab}=R_{\phantom{a}acb}^{c}$ is the Ricci tensor and $R=g^{ab}R_{ab}$
is the Ricci scalar.  In this paper we employ the convention of Wald
\cite{wald1} for the Riemann tensor; the definition is given by equation
(\ref{defriemann}) in Appendix A.  The complete action on a
four-dimensional manifold $(\mathcal{M},g_{ab})$ (assumed for the moment
to have no boundaries) with cosmological constant $\Lambda$ is then given
by
\bea
S = \frac{1}{16\pi G}\int_{\mathcal{M}}d^{4}x\sqrt{-g}
    (R - 2\Lambda + \alpha\mathcal{L}_{GB}) \; .
\label{action1}
\eea
Here, $\alpha$ is the GB parameter.  In four dimensions, the GB term is
a topological invariant of $\mathcal{M}$ known as the Euler
charactersistic $\chi(\mathcal{M})$, and (up to surface terms) does not
contribute to the equations of motion.  In $D\geq5$ dimensions, however,
the GB term is no longer a topological invariant of $\mathcal{M}$ (see
$\S3$ below), and gives non-trivial modifications to the dynamics of
gravity.  This is precisely what happens with the Einstein-Hilbert
action: it is the Euler characteristic of a two-dimensional manifold,
but in $D\geq3$ dimensions describes the dynamics of spacetime!
Therefore the GB term cannot be excluded from the action principle
in dimensions $D\geq5$.  Moreover, from the superstring theory point
of view, the GB term is the only combination of curvature-squared
interactions for which the low-energy effective action is ghost-free
\cite{zwiebach}.  Therefore we consider here the gravitational action
\bea
S = \frac{1}{2k_{D}}\int_{\mathcal{M}}d^{D}x\sqrt{-g}
    (R - 2\Lambda + \alpha\mathcal{L}_{GB}) \, ,
\label{action2}
\eea
where now the indices run $a,b,\ldots\in\{0,\ldots,D-1\}$.  In this
work, we use the standard convention for the coupling constant such
that $k_{D}=8\pi G_{D}$ with $G_{D}$ the $D$-dimensional Newton constant
\cite{myeper}.  The cosmological constant is given by
\bea
\Lambda = \frac{\varepsilon}{2l^{2}}(D-1)(D-2) \, ,
\eea
where $\varepsilon\in\{-1,1\}$ and $l$ is the de Sitter radius \cite{ssv}.
In braneworld models one usually only considers the case for which
$\varepsilon=-1$, i.e. asymptotically adS spacetime.

The equations of motion are given by $\delta S=0$, where $\delta$ is the
first variation; i.e. the stationary points of the action.  Varying the action
(\ref{action2}) with respect to the metric gives the EGB field equations
\bea
G_{ab} &=& -\Lambda g_{ab}
         + \alpha\left[\frac{1}{2}\mathcal{L}_{GB}g_{ab} - 2RR_{ab}
         + 4R_{ac}R_{b}^{\phantom{a}c} + 4R_{acbd}R^{cd}
         - 2R_{acde}R_{b}^{\phantom{a}cde}\right]\nonumber\\
G_{ab} &\equiv& R_{ab} - \frac{1}{2}Rg_{ab} \; .
\label{eom1}
\eea
When $\alpha=0$ these equations reduce to the Einstein field equations
$G_{ab}=-\Lambda g_{ab}$.  The EGB equations admit the following class of
(static) black hole solutions \cite{cai}:
\bea
ds^{2} &=& -h(r)dt^{2} + \frac{dr^{2}}{h(r)} + r^{2}d\Omega_{(k)D-2}^{2}\nonumber\\
h(r) &=& k + \frac{r^{2}}{2\tilde{\alpha}}\left(1 - \sqrt{1
             - \frac{8\tilde{\alpha}\Lambda}{(D-1)(D-2)}
             + \frac{8k_{D}\tilde{\alpha}M}{(D-2)\mathcal{V}_{(k)D-2}r^{D-1}}}\right) \; .
\label{bhsolution}
\eea
Here, $\mathcal{V}_{(k)N-1}=\pi^{N/2}/\Gamma(N/2+1)$ is the volume of an
$(N-1)$-dimensional space $S^{N-1}\equiv S_{(k)}^{N-1}$ of constant curvature with
metric $d\Omega_{(k)N-1}^{2}$; $k$ is the curvature index with $k=1$ corresponding to
positive constant curvature, $k=-1$ corresponding to negative constant curvature, and
$k=0$ corresponding to zero curvature.  $M$ is the mass of the black hole, and
$\tilde{\alpha}$ is related to the GB parameter via
\bea
\tilde{\alpha} = (D-3)(D-4)\alpha \; .
\eea
The singular surfaces with radii $r_{*}$ are given by the roots to the
equation $h(r=r_{*})=0$.  We denote the event horizon by $r_{+}$.  The
location of this surface depends on the sign of the cosmological constant:
if $\Lambda\leq0$ then the largest root $r_{+}$ is the event horizon, and
if $\Lambda>0$ then the largest root is the cosmological horizon and
therefore the second largest root is the event horizon.

The thermodynamics of the black hole is determined in the usual way
\cite{hawking2}.  In particular, the average energy $\ave{E}$ and entropy
$S$ are given by
\bea
\ave{E} = -\frac{\partial}{\partial\beta}(\ln\mathcal{Z})
\quad
\mbox{and}
\quad
S = \beta\ave{E} + \ln\mathcal{Z} \, ,
\label{eands}
\eea
where $\ln\mathcal{Z}$ is the (zero-loop) partition function and
$\beta$ is the inverse temperature.  The partition function is determined
via $\ln\mathcal{Z}=-\tilde{I}[g]$ by evaluating the Euclidean action
$\tilde{I}[g]$ (in the stationary phase approximation where $g$ are
solutions to the equations of motion $\delta\smallint\tilde{I}=0$), and the
inverse temperature is determined by requiring that the Euclidean manifold
does not contain any conical singularities at $r_{+}$ where the manifold
closes up.  For the black hole solution (\ref{bhsolution}) one finds that
\cite{cai}
\bea
\ave{E} = M
\quad
\mbox{and}
\quad
S = \frac{\mathcal{A}_{D-2}r_{+}^{D-2}}{4G_{D}}
    \left[1 + \left(\frac{D-2}{D-4}\right)
    \frac{2\tilde{\alpha}k}{r_{+}^{2}}\right] \; .
\label{entropy1}
\eea
Here, $\mathcal{A}_{N-1}=2\pi^{N/2}/\Gamma(N/2)$ is the surface
area of a unit $(N-1)$-sphere.  This shows that the entropy acquires
a correction due to the presence of the GB term.  A more geometrical
expression for the entropy can be obtained by using the Noether charge
formalism \cite{wald2,iyewal,jkm}.  For EGB gravity, one finds that the
entropy is \cite{crs}
\bea
S = \frac{1}{4G_{D}}\int_{S^{D-2}}d^{D-2}x\sqrt{h}(1 + 2\alpha\mathcal{R}) \, ,
\label{entropy2}
\eea
where $\mathcal{R}=\mathcal{R}_{ij}h^{ij}$ ($i,j,\ldots\in\{0,\ldots,D-2\}$)
is the Ricci scalar determined by the metric $h_{ij}=r_{+}^{2}d\Omega_{(k)D-2}^{2}$
on the surface $S^{D-2}$. Note, however, that this surface need not be a space
of constant curvature.  The assumption in the Noether charge approach is stationarity;
the existence of a globally-defined Killing vector field is required.  One purpose
of the isolated horizon framework is to relax this assumption, and to derive the zeroth
and first laws of black-hole mechanics with minimal conditions assumed about the spacetimes
in question.  In this sense the isolated horizon framework generalizes the notion of a
Killing horizon to include situations where fields outside the horizon may be dynamical.

Let us now proceed to the connection formulation of EGB gravity, which will
pave the way to the construction of the corresponding covariant phase space.

\section{Gauss-Bonnet term in the first-order formulation}

In the connection formulation of general relativity, the configuration space
consists of the pair $(e^{I},A_{\phantom{a}J}^{I})$, where the co-frame
$e^{I}=e_{a}^{\phantom{a}I}dx^{a}$ determines the metric
\bea
g_{ab} = \eta_{IJ}e_{a}^{\phantom{a}I} \otimes e_{b}^{\phantom{a}J} \, ,
\eea
and the connection $A_{\phantom{a}J}^{I}=A_{a\phantom{a}J}^{\phantom{a}I}dx^{a}$
determines the curvature two-form
\bea
\Omega_{\phantom{a}J}^{I}
= dA_{\phantom{a}J}^{I} + A_{\phantom{a}K}^{I} \wedge A_{\phantom{a}J}^{K} \; .
\eea
Internal indices $I,J,\ldots\in\{0,\ldots,D-1\}$ are raised and lowered
using the Minkowski metric $\eta_{IJ}=\mbox{diag}(-1,1,\ldots,1)$.  The curvature
defines the Riemann tensor $R_{\phantom{a}JKL}^{I}$ via
\bea
\Omega_{\phantom{a}J}^{I}
= \frac{1}{2}R_{\phantom{a}JKL}^{I}e^{K} \wedge e^{L} \; .
\eea
The Ricci tensor is then $R_{IJ}=R_{\phantom{a}IKJ}^{K}$, and the
Ricci scalar is $R=\eta^{IJ}R_{IJ}$.  The gauge covariant derivative
$\mathscr{D}$ acts on generic fields $\Psi_{IJ}$ such that
\bea
\mathscr{D}\Psi_{\phantom{a}J}^{I}
= d\Psi_{\phantom{a}J}^{I}
  + A_{\phantom{a}K}^{I} \wedge \Psi_{\phantom{a}J}^{K}
  - A_{\phantom{a}J}^{K} \wedge \Psi_{\phantom{a}K}^{I} \; .
\eea
Finally, the co-frame defines the $(D-m)$-form
\bea
\Sigma_{I_{1}\ldots I_{m}}
= \frac{1}{(D-m)!}\epsilon_{I_{1} \ldots I_{m}I_{m+1} \ldots I_{D}}
e^{I_{m+1}} \wedge \cdots \wedge e^{I_{D}} \, ,
\eea
where the totally antisymmetric Levi-Civita tensor $\epsilon_{I_{1} \ldots I_{D}}$
is related to the spacetime volume element by
\bea
\epsilon_{a_{1} \ldots a_{D}}=\epsilon_{I_{1} \ldots I_{D}}
e_{a_{1}}^{\phantom{a}I_{1}} \cdots e_{a_{D}}^{\phantom{a}I_{D}} \; .
\eea
In this configuration space the action for EGB gravity becomes
\cite{myers,padilla}
\bea
S = \frac{1}{2k_{D}}\int_{\mathcal{M}}\Sigma_{IJ} \wedge \Omega^{IJ}
    - 2\Lambda\bm{\epsilon}
    + \alpha\Sigma_{IJKL} \wedge \Omega^{IJ} \wedge \Omega^{KL} \, ,
\label{action3}
\eea
where $\bm{\epsilon}=e^{0} \wedge \cdots \wedge e^{D-1}$ is the spacetime
volume element.  Here the equations of motion are derived from independently
varying the action with respect to the connection and co-frame.  The equation
of motion for the connection is
\bea
\mathscr{D}(\Sigma_{IJ} + 2\alpha\Sigma_{IJKL} \wedge \Omega^{KL}) = 0 \; .
\label{eom2}
\eea
This equation says that, in general, there exists a non-vanishing torsion
$T^{I}=\mathscr{D}e^{I}$.  To see what constraints are imposed on $T$, we
can use the Bianchi identity $\mathscr{D}\Omega^{IJ}=0$ together with the
identity
\bea
\mathscr{D}\Sigma_{I_{1} \ldots I_{m}}
= \mathscr{D}e^{M} \wedge \Sigma_{I_{1} \ldots I_{m}M} \; .
\eea
Substituting these into equation (\ref{eom2}) gives
\bea
T^{I} \wedge (\Sigma_{IJK} + 2\alpha\Sigma_{IJKLM} \wedge \Omega^{LM}) = 0 \; .
\label{constraint}
\eea
In analogy with Einstein gravity, we assume directly that the torsion in
(\ref{constraint}) vanishes.  (The torsion in Einstein gravity is zero, but
this is not an assumption.  The condition follows directly from the equation
of motion for the connection.)  To get the equation of motion for the co-frame
we note that the variation of $\Sigma$ is given by
\bea
\delta\Sigma_{I_{1} \ldots I_{m}}
= \delta e^{M} \wedge \Sigma_{I_{1} \ldots I_{m}M} \; .
\eea
This leads to
\bea
\Sigma_{IJK} \wedge \Omega^{JK} - 2\Lambda\Sigma_{I}
+ \alpha\Sigma_{IJKLM} \wedge \Omega^{JK} \wedge \Omega^{LM} = 0 \; .
\label{eom3}
\eea
The equations (\ref{eom2}) and (\ref{eom3}) for the connection and co-frame
are equivalent to the equations (\ref{eom1}) in the metric formulation.

\section{Boundary conditions}

The reasons to consider a quasilocal description of black holes were outlined in
$\S1$.  Therefore we proceed directly to the main definitions for the existence
of an isolated horizon, adapted here to EGB gravity in arbitrary dimensions.

We consider a $D$-dimensional spacetime manifold $\mathcal{M}$ with topology
$R\times M$ containing a $(D-1)$-dimensional null surface $\Delta$ as inner
boundary (representing the event horizon), and is bounded by $(D-1)$-dimensional
manifolds $M^{\pm}$ that intersect $\Delta$ in $(D-2)$-spaces $S^{\pm}$ and
extend to the boundary at infinity $\mathscr{B}$.  The manifold $\mathcal{M}$
is said to be globally hyperbolic if it can be foliated by a one-parameter
family of spacelike hypersurfaces $M_{t}$.  It follows that $M_{t}$ are
(partial) Cauchy surfaces.  Then, any wave equation with solutions restricted
to $M$ will have a well defined initial-value formulation (see e.g. \cite{wald3}).
The outer boundary $\mathscr{B}$ is some arbitrary $(D-1)$-dimensional surface,
and is loosely referred to as the ``boundary at infinity''.  In other words, we
consider the purely quasilocal case and neglect any subleties that are associated
with the outer boundary.  See Figure $1$.
\begin{figure}[t]
\begin{center}
\psfrag{D}{$\Delta$}
\psfrag{Mp}{$M^{+}$}
\psfrag{Mm}{$M^{-}$}
\psfrag{Mi}{$M$}
\psfrag{Sp}{$S^{+}$}
\psfrag{Sm}{$S^{-}$}
\psfrag{S}{$S$}
\psfrag{B}{$\mathscr{B}$}
\psfrag{M}{$\mathcal{M}$}
\includegraphics[width=3.8in]{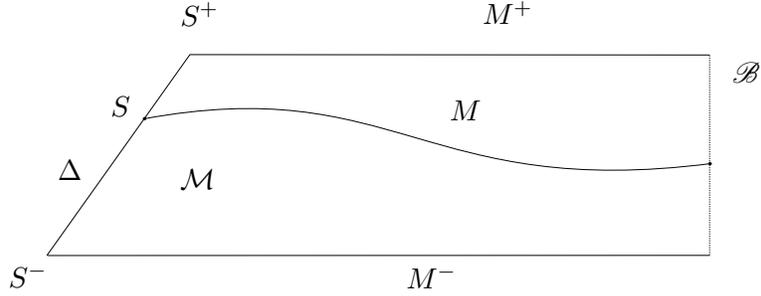}
\caption{The region of the $D$-dimensional spacetime $\mathcal{M}$
being considered has an internal boundary $\Delta$ representing
the event horizon, and is bounded by two (partial) Cauchy surfaces
$M^{\pm}$ which intersect $\Delta$ in $(D-2)$-spaces $S^{\pm}$ and
extend to the boundary at infinity $\mathscr{B}$.}
\end{center}
\end{figure}

\noindent{\bf Definition I.}
\emph{A non-expanding horizon $(\Delta,q_{ab},\ell_{a})$ is a $(D-1)$-dimensional
null hypersurface $\Delta$ (with topology $R\times S^{D-2}$ ) together with a
degenerate metric $q_{ab}$ of signature $0+\ldots+$ (with $D-2$ non-degenerate
spatial directions) and a null normal $\ell_{a}$ such that: (a) the expansion
$\theta_{(\ell)}$ of $\ell_{a}$ vanishes on $\Delta$; (b) the field equations
hold on $\Delta$; and (c) the Ricci tensor is such that
$-R_{\phantom{a}b}^{a}\ell^{b}$ is a future-directed causal vector.}

Condition (c) is analogous to the dominant energy condition imposed on any matter
fields that may be present in the neighbourhood of the horizon; in Einstein gravity
the condition is imposed on the stress-energy tensor, but here the condition must
be imposed directly on the Ricci tensor because $G_{ab}\neq k_{D}T_{ab}$.  Conditions
(a) and (c) hold for any null normal regardless of the normalization of $\ell$.
Condition (a) implies that the surface $\Delta$ is ``time-independent'' in the sense
that all of its cross-sections have the same area.  Condition (a) also implies that
$\Delta$ is a congruence of null geodesics, which in turn implies (by the Frobenius
theorem) that the rotation tensor is zero.  The Raychaudhuri equation then implies
that $R_{ab}\ell^{a}\ell^{b}=-\sigma_{ab}\sigma^{ab}$, where $\sigma_{ab}$ is the
shear tensor.  From condition (c) it follows that $\sigma_{ab}=0$ and
$R_{ab}\ell^{a}\ell^{b}=0$.

The vanishing of the expansion, rotation and shear implies that
$\nabla_{\!\underleftarrow{a}}\ell_{b}\approx\omega_{a}\ell_{b}$.  (We are
using the convention from constrained Hamiltonian systems whereby ``$\approx$''
denotes equality restricted to a submanifold -- in the present context the
restriction is to $\Delta\subset\mathcal{M}$.  The underarrow indicates pull-back
to $\Delta$.)  Thus the one-form $\omega$ is the natural connection (in the normal
bundle) induced on the horizon.  The ``time-independence'' of $\omega$ on $\Delta$
ensures the weak isolation of a non-expanding horizon:

\noindent{\bf Definition II.}
\emph{A weakly isolated horizon $(\Delta,q_{ab},[\ell])$ is a non-expanding horizon
$\Delta$ together with an equivalence class of null normals $[\ell]$ such that
$\pounds_{\ell}\omega_{a}=0$ for all $\ell\in[\ell]$ (where $\ell^{\prime}\sim\ell$
if $\ell^{\prime}=c\ell$ for some constant $c$).}

The above condition is a restriction on the rescaling freedom of $\ell$.  Now, for
any vector $t^{a}$ tangent to $\Delta$ we have that
\bea
t^{a}\nabla_{\! a}\ell^{b} = t^{a}\omega_{a}\ell^{b} \, ;
\eea
in particular, because $\ell^{a}$ is tangent to $\Delta$ we have that
\bea
\ell^{a}\nabla_{\! a}\ell^{b} = \ell^{a}\omega_{a}\ell^{b} \; ,
\eea
which means that $\ell^{a}$ is geodesic.  This defines the surface gravity
$\kappa_{(\ell)}=\ell^{a}\omega_{a}$.  It is important to keep in mind, however,
that $\kappa_{(\ell)}$ is an intrinsic property not of the horizon but of the
null normal; the rescaling freedom of $\ell$ means that if $\ell^{\prime}=f\ell$
for some function $f$, then
$\omega^{\prime}_{a}=\omega_{a}+\nabla_{\!\underleftarrow{a}}f$ and
$\kappa_{(f\ell)}=f\kappa_{(\ell)}+\pounds_{\ell}f$.  Note that under this
rescaling $\omega$ transforms as a connection.  This suggests that
$\kappa_{(\ell)}$ may not be constant on $\Delta$.  It turns out that
$\pounds_{\ell}\omega\approx0$ is sufficient to obtain $d(\ell^{a}\omega_{a})=0$
(see \cite{afk}).  The zeroth law therefore follows from the boundary conditions
and is independent of the functional content of the Lagrangian.

In this paper, for simplicity, we will restrict our attention to \emph{non-rotating}
weakly isolated horizons. That is, we will assume that
$\omega_{a}=-\kappa_{(\ell)}n_{a}$.  Such horizons include, but are not restricted
to, those with spherical symmetry.  The name arises from that fact that the non-$n$
components of $\omega_{a}$ are associated with the angular momentum of a horizon.
Specifically, given a foliation of $\Delta$ into spacelike $(D-2)$-surfaces $S_v$
and a rotational  vector field $\phi^{a}$ parallel to those surfaces, the angular
momentum of the horizon associated with $\phi^{a}$ on a given slice is 
\bea
J_\phi = \oint_{S_{v}}\bm{\tilde{\epsilon}}\phi^{a}\omega_{a} \, , 
\eea
where $\bm{\tilde{\epsilon}}$ is the area form on the surface. Thus, for a
non-rotating horizon, $J_{\phi}=0$ for all rotational vector fields.  For further
discussion of rotational vectors and angular momentum see, e.g. \cite{abl1,boofai2}
or one of the review articles \cite{ashkri,booth,goujar}.

\section{Variation of the boundary term}

We have seen that the boundary conditions for an isolated horizon need to be
modified for EGB gravity by imposing the analogue of the dominant energy
condition directly on the Ricci tensor.  In the action principle, the main
modification to the formalism is the appearance of an additional surface term.
Let us therefore reconsider the action (\ref{action3}) but for a region of the
manifold $\mathcal{M}$ that is bounded by null surface $\Delta$ and spacelike
surfaces $M^{\pm}$ which extend to the (arbitrary) boundary $\mathscr{B}$
(Figure $1$).

Denoting the pair $(e,A)$ collectively as a generic field variable $\Psi$, the
first variation gives
\bea
\delta S = \frac{1}{2k_{D}}\int_{\mathcal{M}}E[\Psi]\delta\Psi
           + \frac{\mathfrak{D}}{2k_{D}}\int_{\partial\mathcal{M}}J[\Psi,\delta\Psi] \; .
\label{first}
\eea
Here $E[\Psi]=0$ symbolically denotes the equations of motion and
\bea
J[\Psi,\delta\Psi] = \widetilde{\Sigma}_{IJ} \wedge \delta A^{IJ}
\label{surface}
\eea
is the surface term, with $\mathfrak{D}=(-1)^{-(D-2)}\equiv(-1)^{D}$ and $(D-2)$-form
\bea
\widetilde{\Sigma}_{IJ} = \Sigma_{IJ} + 2\alpha\Sigma_{IJKL} \wedge \Omega^{KL} \; .
\label{sigmawidetilde}
\eea
If the integral of $J$ on the boundary $\partial\mathcal{M}$ vanishes then the
action principle is said to be differentiable.  We must show that this is the
case.  Because the fields are held fixed at $M^{\pm}$ and at $\mathscr{B}$, $J$
vanishes there.  So we only need to show that $J$ vanishes at the inner boundary
$\Delta$.  To show that this is true we need to find an expression for $J$ in
terms of $A$ and $\widetilde{\Sigma}$ pulled back to $\Delta$.  This is
accomplished by fixing an internal basis consisting of the (null) pair
$(\ell,n)$ and $D-2$ spacelike vectors $\vartheta_{(i)}$
($i\in\{2,\ldots,D-1\}$)
such that
\bea
e_{0} = \ell \, ,
\quad
e_{1} = n \, ,
\quad
e_{i} = \vartheta_{(i)} \, ,
\label{npbasis}
\eea
together with the conditions
\bea
\ell \cdot n=-1 \, ,
\quad
\ell \cdot \ell = n \cdot n = \ell \cdot \vartheta_{(i)} = n \cdot \vartheta_{(i)} = 0 \, ,
\quad
\vartheta_{(i)} \cdot \vartheta_{(j)} = \delta_{ij} \; .
\label{npconditions}
\eea
In the following we also apply the summation convention over repeated
spacelike indices ($i,j,k$ etc.). As these are Euclidean indices their
position (up or down) will be adjusted according to the dictates of
notational convenience.  Thus, we employ a higher-dimensional analogue
of the Newman-Penrose (NP) formalism \cite{ppcm,opp}.

To find the pull-back of $A$ we first note that
\bea
\nabla_{\!\underleftarrow{a}}\ell_{I}
&\approx& \nabla_{\!\underleftarrow{a}}e_{\phantom{a}I}^{b}\ell_{b}\nonumber\\
&\approx& (\nabla_{\!\underleftarrow{a}}e_{\phantom{a}I}^{b})\ell_{b}
              + e_{\phantom{a}I}^{b}\nabla_{\!\underleftarrow{a}}\ell_{b}\nonumber\\
&\approx& e_{\phantom{a}I}^{b}\omega_{a}\ell_{b}\nonumber\\
&\approx& \omega_{a}\ell_{I} \, ,
\label{pullback}
\eea
where we used $\nabla_{\! a}e_{\phantom{a}I}^{b}=0$ in going from the
second to the third line (a consequence of the metric compatibility of
the connection).  Now, taking the covariant derivative of $\ell$ acting
on internal indices gives
\bea
\nabla_{\! a}\ell_{I} = \partial_{a}\ell_{I} + A_{aIJ}\ell^{J} \, ,
\eea
where $\partial$ is a flat derivative operator that is compatible
with the internal co-frame on $\Delta$.  Thus $\partial_{a}\ell_{I}\approx0$
and $\nabla_{\!\underleftarrow{a}}\ell_{I}\approx
A_{\underleftarrow{a}IJ}\ell^{J}$.  Putting this together with
(\ref{pullback}) we have that
$A_{\underleftarrow{a}IJ}\ell^{J}\approx\omega_{a}\ell_{I}$, and this
implies that the pull-back of $A$ to the horizon is of the form
\bea
A_{\underleftarrow{a}}^{\phantom{a}IJ}
\approx -2\ell^{[I}n^{J]}\omega_{a}
+ a_{a}^{(i)}\ell^{[I}\vartheta_{(i)}^{\phantom{a}J]}
+ b_a^{(ij)}\vartheta_{(i)}^{\phantom{a}[I}\vartheta_{(j)}^{\phantom{a}J]} \, ,
\label{pullbackofa}
\eea
where the $a_{a}^{(i)}$ and $b_{a}^{(ij)}$ are one-forms in the cotangent
space $T^{*}(\Delta)$.  It follows that the variation of (\ref{pullbackofa})
is
\bea
\delta A_{\underleftarrow{a}}^{\phantom{a}IJ}
\approx -2\ell^{[I}n^{J]}\delta \omega_{a}
+ \delta a_{a}^{(i)}\ell^{[I}\vartheta_{(i)}^{\phantom{a}J]}
+ \delta b_a^{(ij)} \vartheta_{(i)}^{\phantom{a}[I}\vartheta_{(j)}^{J]} \; .
\label{variationofpullbackofa}
\eea

Then, either by direct calculation from (\ref{pullbackofa}) or from the
considerations of Appendix A, it can be shown that on any weakly isolated
and non-rotating horizon the pull-back of the associated curvature is
\bea
\Omega_{\underleftarrow{ab}}^{\phantom{aa}IJ}
\approx \vartheta^{(k)}_{a}\vartheta^{(l)}_{b}\mathcal{R}_{kl}^{\phantom{aa}ij}
        \vartheta_{(i)}^{\phantom{a}I}\vartheta_{(j)}^{\phantom{a}J}
        +2\ell^{[I}\vartheta_{(i)}^{\phantom{a}J]}\Omega_{\underleftarrow{ab}}^{\phantom{aa}KL}
        \vartheta_{\phantom{a}K}^{(i)} n_L \, ,
\label{curvature}
\eea
where $\mathcal{R}_{kl}^{\phantom{aa}ij}$ is the Riemann tensor associated
with the $(D-2)$ metric $\tilde{q}_{ab} = g_{ab} + \ell_a n_b + n_a \ell_b$. 
That is, given a foliation of $\Delta$ into spacelike $(D-2)$-surfaces, the
spacelike $\vartheta_{(i)}^{a}$ give an orthonormal basis on those surfaces and 
$\mathcal{R}_{kl}^{\phantom{aa}ij} $ is the corresponding curvature tensor; 
for a non-expanding horizon, these quantities are independent of both the 
slice of the foliation and the particular foliation itself. 

To find the pull-back to $\Delta$ of $\widetilde{\Sigma}$, we use the
decomposition
\bea
e_{\underleftarrow{a}}^{\phantom{a}I}
\approx -\ell^{I}n_{a} + \vartheta_{(i)}^{\phantom{a}I}\vartheta^{(i)}_{a} \, ,  
\eea
whence the $(D-2)$-form
\bea
\underleftarrow{\Sigma}_{IJ} &\approx&
-\frac{1}{(D-3)!}
\epsilon_{IJA_1 \dots A_{D-2}} \ell^{A_1} 
\vartheta^{\phantom{a}A_2}_{(i_1)} \dots \vartheta^{\phantom{a}A_{D-2}}_{(i_{D-3})}
\left(
n \wedge \vartheta^{(i_1)} \wedge \dots \wedge \vartheta^{(i_{D-3})} \right) 
\nonumber \\
& &
+ \frac{1}{(D-2)!} \epsilon_{IJA_1 \dots A_{D-2}}  
\vartheta^{\phantom{a}A_1}_{(i_1)} \dots \vartheta^{\phantom{a}A_{D-2}}_{(i_{D-2})}
\left( \vartheta^{(i_1)} \wedge \dots \wedge \vartheta^{(i_{D-2})} \right) \, ,
\label{pullbackofsigmaij}
\eea
and in $D\geq5$ dimensions, the $(D-4)$-form
\bea
\underleftarrow{\Sigma}_{IJKL}
&\approx&
-\frac{1}{(D-5)!} 
\epsilon_{IJKLA_1 \dots A_{D-4}} \ell^{A_1} 
\vartheta^{\phantom{a}A_2}_{(i_1)} \dots \vartheta^{\phantom{a}A_{D-4}}_{(i_{D-5})}
\left(
n \wedge \vartheta^{(i_1)} \wedge \dots \wedge \vartheta^{(i_{D-5})} \right) 
\nonumber \\
& & 
+\frac{1}{(D-4)!} 
\epsilon_{IJKLA_1 \dots A_{D-4}}  
\vartheta^{\phantom{a}A_1}_{(i_1)} \dots \vartheta^{\phantom{a}A_{D-4}}_{(i_{D-4})}
\left( \vartheta^{(i_1)} \wedge \dots \wedge \vartheta^{(i_{D-4})} \right) \; .
\label{pullbackofsigmaijkl}
\eea
In four dimensions $\underleftarrow{\Sigma}_{IJKL} = \epsilon_{IJKL}$.

These expressions are somewhat formidable but on combining them to
find $\widetilde{\Sigma}_{IJ} \wedge \delta A^{IJ}$ there is significant simplification. 
The key is to note that each term includes a total contraction of 
$\epsilon_{I_1 \dots I_{D}}$. This contraction must 
include one copy of each of $\ell^I$, $n^I$, and the $\vartheta^{\phantom{a}I}_{(i)}$
-- else that term will be zero. Similarly the resulting $(D-1)$ form must be proportional
to $n \wedge \vartheta^{(2)} \wedge \dots \wedge \vartheta^{(D-1)}$. Then (\ref{surface})
becomes
\bea
J[\Psi,\delta\Psi]
\approx \bm{\tilde{\epsilon}} \wedge \delta \omega 
& + & \frac{2\alpha}{(D-4)!}\left(  \epsilon_{IJKLA_1\dots A_{D-4}} \ell^I n^J 
\vartheta^{\phantom{a}K}_{(k)} \vartheta^{\phantom{a}L}_{(l)} 
\vartheta^{\phantom{a}A_1}_{(i_1)} \dots \vartheta^{\phantom{a}A_{D-4}}_{i_{D-4}}\right)\nonumber\\
& &
\times \mathcal{R}_{mn}^{\phantom{mn}kl} \vartheta^{(i_1)}
       \wedge \dots \wedge \vartheta^{(i_{D-4})} \wedge \vartheta^{(m)} \vartheta^{(n)}
       \wedge \delta \omega \; .
\eea
The first and second terms respectively come from the
$\underleftarrow{\Sigma}_{IJ}$ and $\underleftarrow{\Sigma}_{IJKL}$
parts of 
$\underleftarrow{\widetilde{\Sigma}}_{IJ}$ while
\bea
\bm{\tilde{\epsilon}} = \vartheta^{(1)} \wedge \dots \wedge \vartheta^{(D-2)}  
\eea
is an area element and we keep in mind that the horizon is non-rotating so that
$\omega_{a}=-\kappa_{(\ell)}n_{a}$.  The second term therefore also simplifies.
Given that there are only $(D-4)$ elements in the spacelike basis it is reasonably
easy to see that this term sums over cases where $(m,n)$ and $(i,j)$ are the same
set of indices. That is (up to a numerical factor) the second term amounts to
contracting $m$ with $i$ and $n$ with $j$ so that the full surface term reduces to
\bea
J[\Psi,\delta\Psi]
\approx \bm{\tilde{\epsilon}}(1 + 2\alpha\mathcal{R}) \wedge \delta \omega \; . 
\label{simplifiedpullbackofcurrent}
\eea

The final step is to note that $\delta\ell\propto\ell$ for some $\ell$ fixed in
$[\ell]$, and this together with $\pounds_{\ell}\omega=0$ implies that
$\pounds_{\ell}\delta\omega=0$.  However, $\omega$ is held fixed on $M^{\pm}$ which
means that $\delta\omega=0$ on the initial and final cross-sections of $\Delta$
(i.e. on $M^{-}\cap\Delta$ and on $M^{+}\cap\Delta$), and because $\delta\omega$ is
Lie dragged on $\Delta$ it follows that $J\approx0$.  Therefore the surface term
$J|_{\partial\mathcal{M}}=0$ for EGB gravity, and we conclude that the equations
of motion $E[\Psi]=0$ follow from the action principle $\delta S=0$.

\section{Covariant phase space and the first law}

In order to derive the first law we need to find the symplectic structure on
the covariant phase space $\bm{\Gamma}$ consisting of solutions $(e,A)$ to the
EGB field equations on $\mathcal{M}$.  Generally, the antisymmetrized second
variation of the surface term gives the symplectic current, and integrating
over a partial Cauchy surface $M$ gives the symplectic structure (the choice
of $M$ being arbitrary).  Following \cite{afk}, we find that the second variation
of the EGB surface term (\ref{surface}) gives
\bea
J[\Psi,\delta_{1}\Psi,\delta_{2}\Psi]
= \mathfrak{D}\left[\delta_{1}\widetilde{\Sigma}_{IJ} \wedge \delta_{2}A^{IJ}
                - \delta_{2}\widetilde{\Sigma}_{IJ} \wedge \delta_{1}A^{IJ}\right] \, ;
\eea
integrating over $M$ defines the \emph{bulk} symplectic structure
\bea
\bm{\Omega}_{\rm bulk}(\delta_{1},\delta_{2})
= \frac{\mathfrak{D}}{2k_{D}}\int_{M}
  \left[\delta_{1}\widetilde{\Sigma}_{IJ} \wedge \delta_{2}A^{IJ}
   - \delta_{2}\widetilde{\Sigma}_{IJ} \wedge \delta_{1}A^{IJ}\right] \; .
\label{bulksymplectic}
\eea
In addition, we need to find the pull-back of $J$ to $\Delta$ and add the
integral of this term to $\bm{\Omega}_{\rm bulk}$ so that the resulting
symplectic structure on $\bm{\Gamma}$ is conserved.  From
(\ref{simplifiedpullbackofcurrent}) we find that
\bea
\bm{\Omega}_{\rm surface}
\approx \frac{\mathfrak{D}}{k_{D}}\int_{\Delta}
\left[\delta_{1}\left[\bm{\tilde{\epsilon}}(1+2\alpha\mathcal{R})\right]
\wedge \delta_{2}\omega
- \delta_{2}\left[\bm{\tilde{\epsilon}}(1+2\alpha\mathcal{R})\right]
\wedge \delta_{1}\omega\right] \; .
\eea
It turns out that this term is a total derivative.  To see this we define
a potential $\psi$ for the surface gravity such that
\bea
\pounds_{\ell}\psi = \kappa_{(\ell)} \; .
\eea
Taking this into account, and using the Stokes theorem, the total derivative
over $\Delta$ becomes an integral over $S^{D-2}$.  The full symplectic structure
for EGB gravity is therefore
\bea
\bm{\Omega}(\delta_{1},\delta_{2})
&=& \frac{1}{2k_{D}}\int_{M}
    \left[\delta_{1}\widetilde{\Sigma}_{IJ} \wedge \delta_{2}A^{IJ}
    - \delta_{2}\widetilde{\Sigma}_{IJ} \wedge \delta_{1}A^{IJ}\right]\nonumber\\
& & + \frac{1}{k_{D}}\oint_{S^{D-2}}
    \left[\delta_{1}\left[\bm{\tilde{\epsilon}}(1+2\alpha\mathcal{R})\right]
    \wedge \delta_{2}\psi
    - \delta_{2}\left[\bm{\tilde{\epsilon}}(1+2\alpha\mathcal{R})\right]
    \wedge \delta_{1}\psi\right] \, ,\nonumber\\
\label{fullsymplectic}
\eea
where we have absorbed the overall (irrelevant) factor of $\mathfrak{D}$.

We can now proceed to derive the first law.  To do so we need to specify
a time evolution vector field $t^{a}$.  Just as for Killing horizons, this
vector field is required to approach an asymptotic time translation at
infinity, and at the horizon must be a symmetry.  Therefore we can restrict
this vector field to the equivalence class $[\ell]$ of null vectors on the
horizon.  (For a rotating horizon we would also add a rotational vector
$\Omega R^{a}$ with $\Omega$ the angular velocity of the horizon.)  The
system is said to be Hamiltonian iff there exists a function $H_{t}$ such
that
\bea
\bm{\Omega}(\delta,\delta_{t}) = \delta H_{t} \; .
\eea
Evaluating the symplectic structure (\ref{fullsymplectic}) with $(\delta,\delta_{t})$
gives two surface terms, one at infinity (which is identified with the ADM energy),
and one at the horizon.  At the horizon, we find that
\bea
\bm{\Omega}|_{\Delta}(\delta,\delta_{t})
= \frac{\kappa_{(t)}}{k_{D}}\delta\oint_{S^{D-2}}
  \bm{\tilde{\epsilon}}(1+2\alpha\mathcal{R}) \; .
\eea
Here, we used $\kappa_{(t)}=\pounds_{t}\psi=t\cdot\omega$.  The right hand side
will be a total variation if the normalization of $t^{a}$ is chosen such that
the functional dependence of the surface gravity is
$\kappa_{(t)}=\kappa_{(t)}(\oint_{S^{D-2}}
\bm{\tilde{\epsilon}}(1+2\alpha\mathcal{R}))$.  The vector fields with this type
of normalization are commonly referred to as ``live'' vector fields.  For details
see e.g. \cite{afk} for non-rotating horizons and \cite{abl1} for rotating horizons.
With this choice made, the right hand side in the above expression is a total
variation, i.e. there exists a function $E_{\Delta}$ such that
$\bm{\Omega}|_{\Delta}(\delta,\delta_{t})=\delta E_{\Delta}$.  We conclude that
\bea
\delta E_{\Delta} = \frac{\kappa_{(t)}}{k_{D}}\delta\oint_{S^{D-2}}\bm{\tilde{\epsilon}}
                    (1 + 2\alpha\mathcal{R}) \, ,
\label{firstlaw}
\eea
which is the first law for the isolated horizon with energy $E_{\Delta}$.  In its
standard form, the first law of thermodynamics (for a quasi-static process) is
$\delta E=T\delta\mathcal{S}+(\mbox{work terms})$.  Here, the temperature is
$T=\kappa_{(t)}/2\pi$.  This identifies the entropy of the isolated horizon:
\bea
\mathcal{S} = \frac{1}{4G_{D}}\oint_{S^{D-2}}\bm{\tilde{\epsilon}}(1 + 2\alpha\mathcal{R}) \; .
\label{entropy3}
\eea
This expression is in exact agreement with the Noether charge expression
(\ref{entropy2}).  As in that approach, no assumptions about the cross sections
$S^{D-2}$ of the horizon need to be made.  An important difference, however, is
that we did not assume the existence of a globally-defined Killing vector.
Instead we had to specify the existence of a time translation vector field
which mimics the properties of a Killing vector but \emph{is not defined
for the entire spacetime}.  For the black hole solution (\ref{bhsolution})
with $\Lambda=0$ and $k=1$, the Ricci scalar is
$\mathcal{R}=(D-2)(D-3)/r_{+}^{2}$ (the Ricci scalar of a $(D-2)$-sphere
with radius $r_{+}$), and (\ref{entropy3}) reduces to (\ref{entropy1}).
Our entropy expression is therefore in agreement with the Euclidean
expression as well.  In our derivation, however, the entropy
(\ref{entropy3}) automatically satisfies the first law (\ref{firstlaw}).
Note that it is possible to have black holes with negative entropies for
negative constant curvature horizons when $2\alpha\mathcal{R}<1$.  This was
first discovered by Cveti$\check{\mbox{c}}$ \emph{et al} in \cite{cno} and
later confirmed by Clunan \emph{et al} \cite{crs}.  For non-rotating horizons,
the first law (\ref{firstlaw}) implies that the energy is also negative; this
is not surprising, as negative-energy solutions are possible when $\Lambda<0$
\cite{hormye}.

\section{Discussion}

We have shown that the isolated horizon framework can be extended beyond Einstein
gravity.  By constructing a covariant phase space for EGB gravity in arbitrary
dimensions, we derived an expression for the entropy of the corresponding isolated
horizons.  This derivation is classical.  The next step is to study the quantum
geometry of the horizons using the state-counting arguments that were developed
by Ashtekar \emph{et al} \cite{abck,abk,domlew}, specifically for the
five-dimensional solution (\ref{bhsolution}).  This should lead to some
interesting physics.  In fact, inclusion of the GB term has physical effects
in four dimensions as well, because the variation
$\delta\mathcal{L}_{GB}|_{\mathcal{M}_{4}}=\delta\chi(\mathcal{M}_{4})$ gives
a surface term that cannot be excluded if $\mathcal{M}_{4}$ has boundaries
\cite{liko}.

In this paper we considered vacuum gravity.  An obvious question is whether
the first law holds for cases where gravity is coupled to matter.  This has
been studied extensively for Einstein gravity in four dimensions
\cite{ashcor,corsud,afk,acs}.  The situation is different in higher dimensions.
For instance, the only Lagrangian for gravity coupled to electromagnetism in
four dimensions is the EM Lagrangian
\bea
S_{EM} = \frac{1}{16\pi G}\int_{\mathcal{M}}
         \Sigma_{IJ} \wedge \Omega^{IJ}
         - \frac{1}{4}\bm{F} \wedge \star \bm{F} \, ,
\eea
where $\bm{F}=d\bm{A}$ is the curvature of the potential one-form $\bm{A}$ and
$\star$ denotes the Hodge dual.  In $D\geq5$ dimensions, however, one can add
to the action a Chern-Simons (CS) term $\bm{A} \wedge \bm{F}^{n}$ ($n=D/2-1$)
for the Maxwell fields.  Of particular interest is the action in five dimensions,
which describes minimal ($N=1$) supergravity and is known to admit black hole
solutions with non-vanishing Killing spinors \cite{gmt}.

One of the main assumptions that we made in our calculations was that the
horizons are non-rotating.  Extension of the phase space of solutions to
include rotation by relaxing the condition $\tomega=0$ would be of interest,
which can be done by using the framework for rotating horizons that was
developed in \cite{abl1}.

The formalism presented here can be further extended by including torsion.
Recall that in $\S3$ we assumed $T^{I}=0$ directly, which became crucial
when we derived the pull-back to $\Delta$ of the connection.  However, as
the equation of motion for $A$ indicates, the torsion-free condition is not
imposed in $D\geq5$ dimensions; in four dimensions
$\Sigma_{IJKL}=\epsilon_{IJKL}$ so that equation (\ref{eom2}) reduces to
$\mathscr{D}e=0$ by virtue of the Bianchi identity $\mathscr{D}\Omega=0$.
If the torsion is non-zero in $D\geq5$ dimensions then the pull-back to
$\Delta$ of $A$ is not given by (\ref{pullbackofa}).  In order to derive
the modified pull-back of $A$ in the presence of torsion we would need to
find $\nabla_{\! \underleftarrow{a}}e_{\phantom{a}I}^{b}$ explicitly.  In
addition, the Raychaudhuri equation would be different as well, and so the
boundary conditions would require a more careful analysis.  The effects of
torsion on isolated horizons should therefore lead to some interesting
consequences.  This would be a particularly interesting project to work
out in five dimensions, for which a constant-curvature black hole is known
to have an entropy that is proportional to the surface area of the inner
horizon rather than the event horizon \cite{banados}.  To study this
curiosity within the isolated horizon framework would require a modification
of the boundary conditions from horizon topology $R\times S^{3}$ to
$R^{3}\times S^{1}$, which is more or less a dimensional continuation of the
three-dimensional isolated horizons that was developed in \cite{adw}.

\section*{Acknowledgements}

We thank Kirill Krasnov for correspondence.  TL would also like to thank the
participants at BH$6$ and at CCGRRA$12$ for discussions related to this work,
especially Kristin Schleich and Don Witt.  This work was supported by the Natural
Sciences and Engineering Research Council of Canada through a PGS D3 Scholarship
(TL) and a Discovery Grant (IB).

\appendix

\section{Restrictions to the Riemann tensor on $\Delta$}

In this appendix we show that for a weakly isolated and non-rotating horizon
\bea
\Omega_{\underleftarrow{ab}}^{\phantom{aa}IJ}
\approx \vartheta^{(k)}_a \vartheta^{(l)}_b \mathcal{R}_{kl}^{\phantom{aa}ij}
\vartheta_{(i)}^{\phantom{a}I}\vartheta_{(j)}^{\phantom{a}J}
+ 2 \ell^{[I} \vartheta_{(i)}^{\phantom{a}J]}\Omega_{\underleftarrow{ab}}^{\phantom{aa}KL}
\vartheta_{\phantom{a}K}^{(i)} n_{L} \, , 
\eea
as stated in equation (\ref{curvature}). 

First we establish some notation. The pull-back operator onto $\Delta$ may be
written as
\bea
q_{a}^{\phantom{a}b} = g_{a}^{\phantom{a}b} + \ell_{a}n^{b} \, , 
\eea
while the pull-back operator into the tangent subspace spanned by the spacelike
$\theta^{a_{(i)}}$ is 
\bea
\tq_{a}^{\phantom{a}b} = g_{a}^{\phantom{a}b} + \ell_{a}n^{b}
                       + n_{a}\ell^{b}
                       = q_{a}^{\phantom{a}b} + n_{a}\ell^{b}
\label{projtq}
\eea
and $\tq_{ab}$ is also the metric on this subspace. We also define the angular
momentum density 
\bea
\tomega_{a} \equiv \tq_{a}^{\phantom{a}b} \omega_{b} \; .  
\eea
It is clear that 
\bea
\omega_{a} = - \kappa_{(\ell)} n_{a} + \tomega_{a} \, ,
\label{decomp}
\eea
and so a horizon is non-rotating if and only if $\tomega_{a}$ vanishes. 

In thinking about these quantities it is useful to keep in mind the case where
$\Delta$ is foliated into spacelike $(D-2)$-surfaces $S_{v}$ which are labelled
by a parameter $v$ and $n$ is chosen to be $-dv$.  Then $\ell^{a}$ evolves the
foliation surfaces while $\ell^{a}$ and $n^{a}$ together span the normal bundle
$T^{\perp}(S_{v})$ on which $\tomega_{a}$ is the connection.  Furthermore, the
$\vartheta_{(i)}^{a}$ span the tangent bundle $T(S_{v})$ and $\tq_{ab}$ is the
metric tensor for the $S_{v}$.

We now turn to the Riemann tensor with the first two indices pulled back to $\Delta$. 
By definition,
\bea
R_{\underleftarrow{ab}\phantom{a}d}^{\phantom{aa}c}\ell^{d}
= -q_{a}^{\phantom{a}e}q_{b}^{\phantom{a}f}
  (\nabla_{e}\nabla_{f} - \nabla_{f}\nabla_{e})\ell^{c} \, , 
\label{defriemann}
\eea
and with the horizon identity $\nabla_{\underleftarrow{a}}\ell^{b}=\omega_{a}\ell^{b}$
along with the decomposition (\ref{decomp}), a few lines of algebra gives
\bea
R_{\underleftarrow{ab}\phantom{a}d}^{\phantom{aa}c}\ell^{d} 
= \left(-2n_{[a} \tilde{d}_{b]}\kappa_{(\ell)} + 2\tq_{[a}^{\phantom{a}e}\tq_{b]}^{\phantom{a}f}
  \tilde{d}_{e}\tomega_{f}
  - 2n_{[a}\tq_{b]}^{\phantom{a}f}\mathcal{L}_{\ell}\tomega_{f}\right)\ell^{c} \, , 
\eea
where $\tilde{d}_{a}$ is the covariant derivative that is compatible with the metric $\tq_{ab}$.
For a weakly isolated horizon the zeroth law ensures that $\tilde{d}_{a}\kappa_{(\ell)}=0$ and
if the horizon is non-rotating then $\tomega=0$ also, whence
\bea
R_{\underleftarrow{ab}\phantom{a}d}^{\phantom{aa}c}\ell^{d} = 0 \; . 
\label{Rl}
\eea
Finally, using this result and (\ref{projtq}) it is straightforward to see that 
\bea
R_{\underleftarrow{ab}cd} \vartheta^c_{(i)} \vartheta^d_{(j)} 
= \tq_{a}^{\phantom{a}e}\tq_{b}^{\phantom{a}f}R_{efcd}\vartheta^c_{(i)}\vartheta^d_{(j)} \;  . 
\eea
From here one can use the fact that 
\bea
\tilde{d}_{a}\tilde{d}_{b}\vartheta^{(i)}_{c}
= \tq_{a}^{\phantom{a}d}\tq_{b}^{\phantom{a}e} \tq_{c}^{\phantom{a}f}
  \nabla_{d}\left(\tq_{e}^{\phantom{a}g}\tq_{f}^{\phantom{a}h}\nabla_{g}\vartheta^{(i)}_{h}\right) \, , 
\eea
and the identity for the Riemann tensor $\mathcal{R}_{abcd}$ associated with 
$\tq_{ab}$ 
\bea
\mathcal{R}_{abcd} \vartheta^{(i)d}
= \left(\tilde{d}_{a}\tilde{d}_{b} - \tilde{d}_{b}\tilde{d}_{a}\right)\vartheta_c^{(i)} \, , 
\eea
along with (\ref{projtq}) to show the Gauss relation
\bea
\tq_{a}^{\phantom{a}e}\tq_{b}^{\phantom{a}f}\tq_{c}^{\phantom{a}g}\tq_{d}^{\phantom{a}h}R_{efgh} 
= \mathcal{R}_{abcd} + \left(k^{(\ell)}_{ac}k^{(n)}_{bd} + k^{(n)}_{ac}k^{(\ell)}_{bd})
-  (k^{(\ell)}_{bc}k^{(n)}_{ad} + k^{(n)}_{bc}  k^{(\ell)}_{ad}\right) \; . 
\label{Gauss}
\eea
Here $k^{(\ell)}_{ab}=\tq_{a}^{\phantom{a}c}\tq_{b}^{\phantom{a}d}\nabla_{c}\ell_{d}$
and $k^{(b)}_{ab}=\tq_{a}^{\phantom{a}c}\tq_{b}^{\phantom{a}d}\nabla_{c}n_{d}$ are the
extrinsic curvatures associated with $\ell_{a}$ and $n_{a}$.  However,
$k^{(\ell)}_{ab}=(1/2)\theta_{(\ell)}\tq_{ab}+\sigma_{ab}$, and on a non-expanding
horizon both the expansion and shear vanish. Thus for the cases in which we are
interested
\bea
\tq_{a}^{\phantom{a}e}\tq_{b}^{\phantom{a}f}\tq_{c}^{\phantom{a}g}\tq_{d}^{\phantom{a}h}R_{efgh} 
= \mathcal{R}_{abcd} \; .
\label{GaussNEH}
 \eea
 
Then equation (\ref{curvature}) directly follows on expanding the frame indices of 
$\Omega_{\underleftarrow{ab}}^{\phantom{aa}IJ}$ in terms of the $\ell^{I}$, $n^{I}$
and $\vartheta_{(i)}^{\phantom{a}I}$, and applying (\ref{Rl}) and (\ref{GaussNEH}).


\end{document}